\documentclass[a4paper,11pt]{article}
\usepackage{pos}
\usepackage{physics}
\usepackage{array, booktabs}

\title{Relationship between the Euclidean and Lorentzian versions of the type IIB matrix model}
\ShortTitle{Euclidean and Lorentzian versions of type IIB matrix model}

\author*[a]{Kohta Hatakeyama}
\author[b]{Konstantinos Anagnostopoulos}
\author[c]{Takehiro Azuma}
\author[a]{Mitsuaki Hirasawa 
}
\author[d]{Yuta Ito}
\author[a,e]{Jun Nishimura}
\author[b]{Stratos Papadoudis}
\author[f]{Asato Tsuchiya}
\affiliation[a]{Theory Center, Institute of Particle and Nuclear Studies,\\
High Energy Accelerator Research Organization (KEK),\\
1-1 Oho, Tsukuba, Ibaraki 305-0801, Japan}
\affiliation[b]{Physics Department, School of Applied Mathematical and Physical Sciences, National Technical University of Athens, Zografou Campus, \\
GR-15780 Athens, Greece}
\affiliation[c]{Setsunan University,\\
17-8 Ikeda Nakamachi, Neyagawa, Osaka, 572-8508, Japan}
\affiliation[d]{National Institute of Technology, Tokuyama College,\\
Gakuendai, Shunan, Yamaguchi 745-8585, Japan}
\affiliation[e]{Department of Particle and Nuclear Physics, School of High Energy Accelerator Science,\\
Graduate University for Advanced Studies (SOKENDAI),\\
1-1 Oho, Tsukuba, Ibaraki 305-0801, Japan
}
\affiliation[f]{Department of Physics, Shizuoka University,\\
836 Ohya, Suruga-ku, Shizuoka 422-8529, Japan}

\emailAdd{khat@post.kek.jp}
\emailAdd{konstant@mail.ntua.gr}
\emailAdd{azuma@mpg.setsunan.ac.jp}
\emailAdd{mitsuaki@post.kek.jp}
\emailAdd{y-itou@tokuyama.ac.jp}
\emailAdd{jnishi@post.kek.jp}
\emailAdd{sp10018@central.ntua.gr}
\emailAdd{tsuchiya.asato@shizuoka.ac.jp}

\abstract{
The type IIB matrix model was proposed as a non-perturbative formulation of superstring theory in 1996. We simulate a model that describes the late time behavior of the
IIB matrix model by applying the 
complex Langevin method to overcome the sign problem. We clarify the relationship between the Euclidean and the Lorentzian versions of the type IIB matrix model in a recently discovered phase.
By introducing a constraint, we obtain a model where the spacetime metric is Euclidean at early times, whereas it {\it dynamically} becomes Lorentzian at late times.
}

\FullConference{%
 The 38th International Symposium on Lattice Field Theory, LATTICE2021
  26th-30th July, 2021
  Zoom/Gather@Massachusetts Institute of Technology
}

\begin{document}

\begin{flushright}
KEK-TH-2373
\end{flushright}
\maketitle

\section{Introduction}
\label{sec: intro}
Superstring theory is the most promising candidate for a unified theory of all interactions, including quantum gravity.
The appearance of extra dimensions has led to the proposal of their compactification to small, unobservable internal spaces, resulting in a vast string landscape
of equivalent vacua. Therefore, it is interesting to study whether non-perturbative effects will lift this degeneracy and allow us to determine the true vacuum of the theory.
Furthermore, other outstanding problems, such as the resolution of the cosmic singularity \cite{Lawrence:2002aj,Liu:2002kb,Horowitz:2002mw,Berkooz:2002je}, may find their 
solution by properly taking account of non-perturbative effects. The IIB matrix model \cite{Ishibashi:1996xs} has been proposed as a non-perturbative definition of superstring theory
and provides a promising context to study such questions.

The IIB matrix model is formally obtained by the dimensional reduction of ten-dimensional ${\cal N}=1$ Super Yang-Mills (SYM) to zero dimensions. 
The theory has maximal  ${\cal N}=2$ supersymmetry (SUSY), where translations are realized by  the shifts $A_\mu\to A_\mu + \alpha_\mu \mathbf{1}$, $\mu=0,\ldots,9$.
It is possible to interpret the eigenvalues of the bosonic matrices $A_\mu$ as defining spacetime points of the target space and the scenario of the
emergence of spacetime from the dynamics of the theory becomes viable. One may study questions, such as the appearance of time, cosmological expansion, and the 
compactification of extra dimensions, as a dynamical effect of the model. For example, in the Euclidean version of the model, dynamical compactification of extra dimensions
results from the Spontaneous Symmetry Breaking (SSB) of the SO(10) rotational symmetry down to SO(3), giving a three-dimensional macroscopic universe. 
The application of the Gaussian Expansion Method (GEM) \cite{Nishimura:2001sx,Kawai:2002jk,Aoyama:2006rk,Nishimura:2011xy}, and 
Monte Carlo calculations \cite{Anagnostopoulos:2013xga,Anagnostopoulos:2015gua,Anagnostopoulos:2017gos,Anagnostopoulos:2020xai} provide strong evidence in support
of this scenario.

There have been several attempts to study the IIB matrix model via Monte Carlo simulations. The problem is hard due to the appearance of a  strong complex
action problem. In  \cite{Kim:2011cr}, an approximation was used to eliminate it, and it was found that a {\it continuous} time emerges from the dynamics of the model, 
with respect to which the universe is expanding. The expansion is exponential at short times and power--like at late times \cite{Ito:2013qga,Ito:2013ywa,Ito:2015mxa,Ito:2015mem}, and
space has three large dimensions, resulting from the SSB of SO(9) rotational symmetry down to SO(3). Space is noncommutative, but at late times
classical solutions dominate giving smooth space and phenomenologically consistent matter content at low energies 
\cite{Kim:2011ts,Kim:2012mw,Chaney:2015ktw,Stern:2018wud,Steinacker:2010rh,Chatzistavrakidis:2010xi,Chatzistavrakidis:2011gs,Steinacker:2017bhb,Aoki:2010gv,Aoki:2014cya,Honda:2019bdi,
Hatakeyama:2019jyw,Steinacker:2021yxt}. In  \cite{Aoki:2019tby}, however, it was shown that SSB comes from singular configurations associated with the Pauli matrices, in which only two eigenvalues are large.
Therefore, it becomes necessary to study the model without the approximation used in  \cite{Kim:2011cr}. To make the model well-defined, the authors in \cite{Nishimura:2019qal}
proposed a two-parameter deformation of the model, corresponding to two independent Wick rotations, on the worldsheet and target space, respectively.
The parameters are $s$ and $k$, respectively, and the model is defined in the  $s,k\to 0$ limit.

Even the deformed model has a strong complex action problem. In \cite{Nishimura:2019qal}, the Complex Langevin Method (CLM)
\cite{Parisi:1983mgm,Klauder:1983sp} was used successfully. Although the CLM is known to lead to wrong results in some cases, the
application of new techniques and easy--to--compute criteria of correct convergence
\cite{Aarts:2009dg,Aarts:2009uq,Aarts:2011ax,Nishimura:2015pba,Nagata:2015uga,Nagata:2016vkn,Ito:2016efb}, make the method possible
to use in a region of parameter space that was not possible to do before. In particular, the singular drift problem
\cite{Nishimura:2015pba} often appears when one studies the effects of dynamical fermions. Using special deformation techniques
\cite{Ito:2016efb} that shift the eigenvalues of the effective fermionic action away from zero made the application of the CLM
successful, at the expense of introducing a new parameter that must be extrapolated to zero in the end. Such techniques have been
applied successfully to the Euclidean version of the IIB matrix model, which were found to agree with the GEM
\cite{Anagnostopoulos:2017gos,Anagnostopoulos:2020xai}.

In this work we study the bosonic version of the  IIB matrix model using the CLM. The model is obtained by quenching the fermionic degrees of freedom, and it is
a simplified model that describes the late time behavior
of the IIB matrix model cosmology.  The model is Wick-rotated as in \cite{Nishimura:2019qal}, using the parameter $u=s=k$, with $0\leq u \leq 1$. 
When $u=0$, we have the original IIB matrix model, whereas when $u=1$, we obtain the Euclidean version of the IIB matrix model, which is 
equivalent to the one studied in \cite{Anagnostopoulos:2017gos,Anagnostopoulos:2020xai}. We find that the parameter $u$ smoothly interpolates between the two models
and expectation values can be obtained by analytic continuation from one model to the other. The two models are equivalent and spacetime in the Lorentzian model turns out to be
Euclidean. In order to study the possibility of the dynamical change of signature from Euclidean to Lorentzian, we introduce a constraint that breaks the 
equivalence between the two models. We find some evidence that the signature of spacetime, although Euclidean at early times, may turn out to be Lorentzian at later times.

\section{The type IIB matrix model}
\label{sec: IIBMM}

\subsection{Definition}
\label{sec: def_IIBMM}
The action of the type IIB matrix model is given as follows:
\begin{align}
\label{eq: action}
S&= S_\mathrm{b}+S_\mathrm{f}\ , \\
\label{eq: action_b}
S_\mathrm{b}&= -\frac{1}{4g^2} \Tr \qty( [A^\mu,A^\nu][A_\mu,A_\nu]) \ , \\
\label{eq: action_f}
S_\mathrm{f}&= -\frac{1}{2g^2} \Tr \qty(\bar{\Psi}(\mathcal{C}\Gamma^\mu) [A_\mu,\Psi])\ ,
\end{align}
where $A_\mu \ (\mu=0,\ldots,9)$ and $\Psi$ are $N\times N$ Hermitian matrices, and $\Gamma^\mu$  and $\mathcal{C}$ are 10-dimensional gamma matrices and the charge conjugation matrix, 
respectively, which are obtained after the Weyl projection. 
The  $A^\mu$ and $\Psi$ transform as vectors and  Majorana-Weyl spinors under SO(9,1) transformations.
In this study, we omit $S_\mathrm{f}$ to reduce the computational cost. The resulting model is expected to describe the late-time behavior of the matrix model cosmology.

The partition function is given by
\begin{equation}
\label{eq: partition_func}Z=\int dA e^{iS_\mathrm{b}}\ .
\end{equation}
Due to the phase factor $e^{iS_\mathrm{b}}$, the model is not well-defined, and we have to define it via analytic continuation. We perform a double Wick rotation onto the complex plane, 
using two parameters $s$ and $k$. The parameter $s$ corresponds to a Wick rotation on the worldsheet, and $k$ to a Wick rotation in target space:
\begin{align}
\label{k1}
\tilde{S}_\mathrm{b}&=-i N\beta 
e^{i\frac{\pi}{2}s} \qty[-\frac{1}{2}e^{-ik\pi}\Tr(F_{0i})^2 +\frac{1}{4} \Tr(F_{ij})^2]\ ,\\
Z&=\int dA e^{-\tilde{S}_\mathrm{b}} \ ,
\end{align}
where $N\beta=1/g^2$ 
and $F_{\mu\nu}=i\comm{A_\mu}{A_\nu}$.
The model (\ref{eq: partition_func}) is obtained in the $(s,k)\to (0,0)$ limit. $(s,k)=(1,1)$ is the Euclidean version of the IIB matrix model and $(s,k)=(-1,0)$ is the simplified model studied in 
\cite{Kim:2011cr}. 

\subsection{Equivalence between the Euclidean and Lorentzian models}
The Wick--rotated matrices $\tilde A_\mu$ are given be the relations
\begin{align}
\label{eq: A0}
A_0&=e^{i\frac{\pi}{8}s -i\frac{\pi}{2}k}\tilde{A}_0
=e^{-i\frac{3\pi}{8} u}\tilde{A}_0 \ ,\\
\label{eq: Ai}
A_i&=e^{i\frac{\pi}{8} s}\tilde{A}_i =e^{i\frac{\pi}{8} u}\tilde{A}_i\ ,
\end{align}
where we introduce the parameter $u$ by setting $s=k=u$. $u=0$ corresponds to the Lorentzian model and $u=1$ to the Euclidean one.

For $(s,k)=(1,1)$,  the Wick-rotated action is given by
\begin{equation}
\tilde{S} = N \beta 
\qty[\frac{1}{2}\Tr(\tilde F_{0i})^2 +\frac{1}{4} \Tr(\tilde F_{ij})^2]\ ,
\end{equation}
where $\tilde F_{\mu\nu}=i\comm{\tilde{A}_\mu}{\tilde{A}_\nu}$. This is the action of the bosonic Euclidean IIB matrix model.

By using Eqs. \eqref{eq: A0} and \eqref{eq: Ai}, one can derive relationships between the expectation values of $\Tr A_0^2$ and $\Tr A_i^2$ in both models:
\begin{align}
\label{eq: TrA0sq}
\expval{\frac{1}{N} \Tr A_0^2}_\mathrm{L}
&=e^{-i\frac{3\pi}{4}}\expval{\frac{1}{N} \Tr \tilde{A}_0^2}_\mathrm{E}\ ,\\
\label{eq: TrAisq}
\expval{\frac{1}{N} \Tr A_i^2}_\mathrm{L}
&=e^{i\frac{\pi}{4}}\expval{\frac{1}{N} \Tr \tilde{A}_i^2}_\mathrm{E}\ ,
\end{align}
where $\expval{\ \cdot \ }_\mathrm{L}$ and $\expval{\ \cdot \ }_\mathrm{E}$ denote the expectation values in the Lorentzian and Euclidean models, respectively.
\begin{figure}
\centering
\includegraphics[scale=0.5]{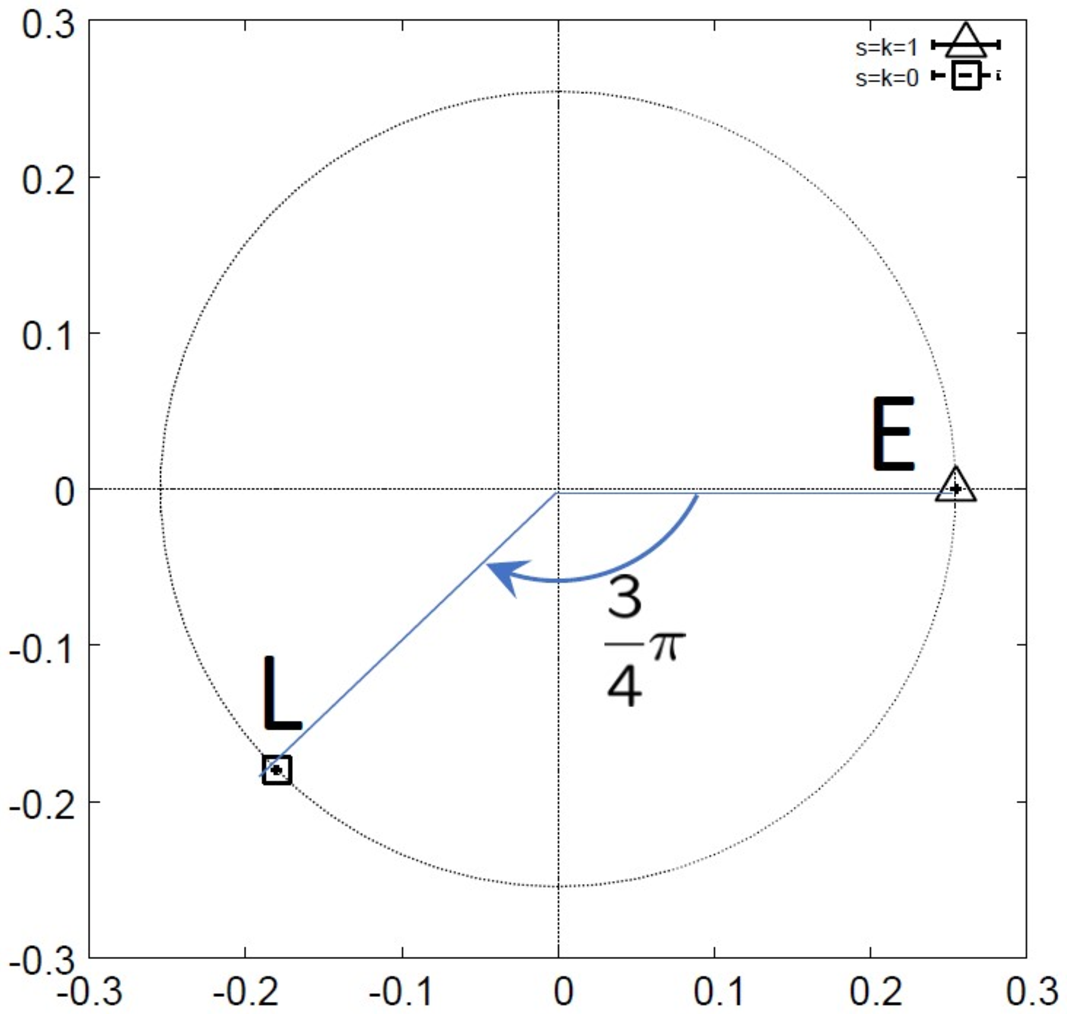} 
\includegraphics[scale=0.5]{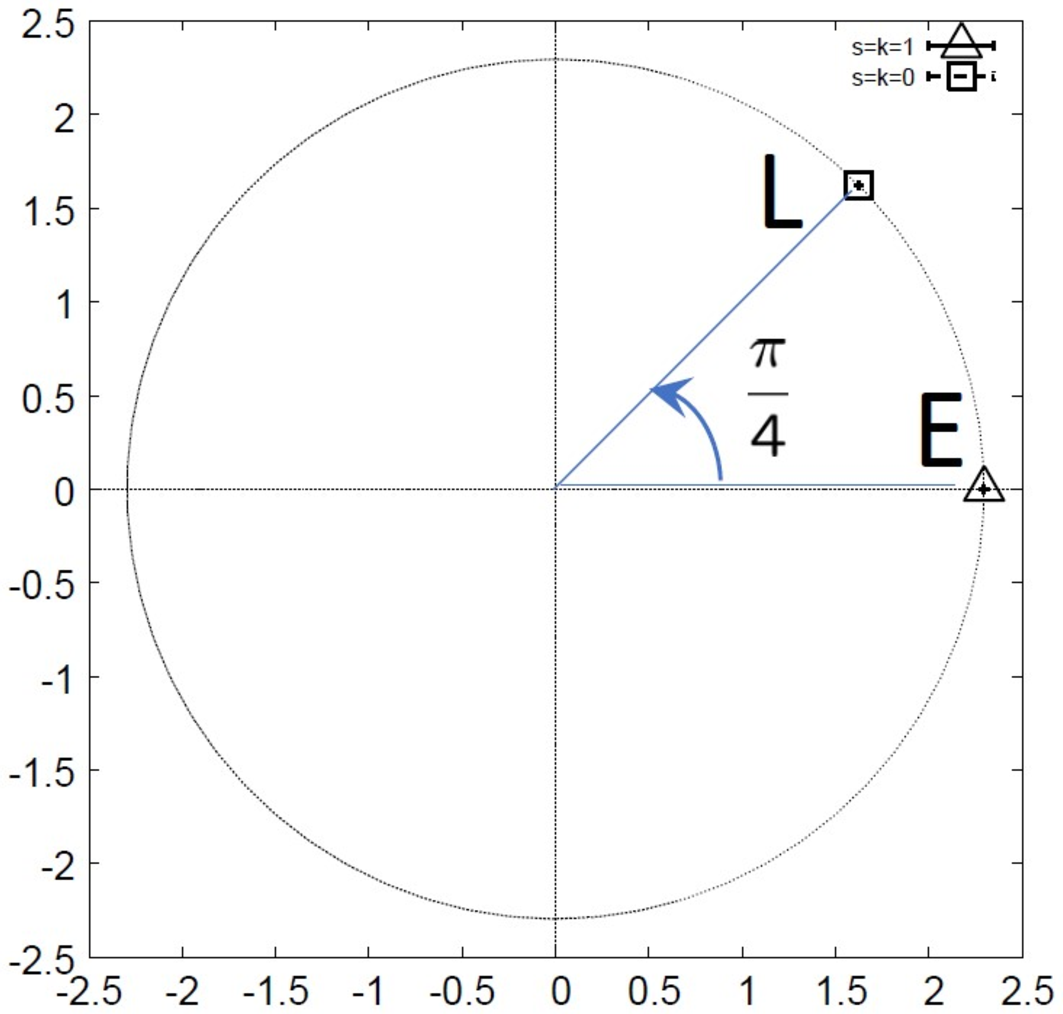} 
\caption{We plot the expectation values of $\frac{1}{N}\Tr A_0^2$ (Left), and $\frac{1}{N}\Tr A_i^2$ (Right). 
Those of the Euclidean model are represented by the triangle and the letter ``{\sf E}'', and those of the Lorentzian model are represented by the square and the letter ``{\sf L}''.
The angle between $\expval{\frac{1}{N} \Tr \tilde{A}_0^2}_\mathrm{E}$ and $\expval{\frac{1}{N} \Tr A_0^2}_\mathrm{L}$ is $-3\pi/4$, in agreement with Eq. \eqref{eq: TrA0sq}, and
the angle between $\expval{\frac{1}{N} \Tr \tilde{A}_i^2}_\mathrm{E}$ and $\expval{\frac{1}{N} \Tr A_i^2}_\mathrm{L}$ is $\pi/4$, in agreement with Eq. \eqref{eq: TrAisq}.}
\label{fig: TrAisq}
\end{figure}
In Fig. \ref{fig: TrAisq} (Left), $\expval{\frac{1}{N}\Tr A_0^2}$ is shown, and the angle between  $\expval{\frac{1}{N} \Tr \tilde{A}_0^2}_\mathrm{E}$ and $\expval{\frac{1}{N} \Tr A_0^2}_\mathrm{L}$ is $-3\pi/4$.
In Fig. \ref{fig: TrAisq} (Right), $\expval{\frac{1}{N}\Tr A_i^2}$ is shown, and the angle between  $\expval{\frac{1}{N} \Tr \tilde{A}_i^2}_\mathrm{E}$ and $\expval{\frac{1}{N} \Tr A_i^2}_\mathrm{L}$ is $\pi/4$.
These angles are in agreement with Eqs. \eqref{eq: TrA0sq} and \eqref{eq: TrAisq}.

These results are consistent with the fact that the Lorentzian and the Euclidean models are equivalent. Expectation values in the Lorentzian model can be obtained by analytic continuation of the
expectation values in the Euclidean model. In particular, $\expval{\frac{1}{N} \Tr A_\mu^2}_\mathrm{L}$ are complex and the emergent spacetime should be interpreted to be Euclidean.

\subsection{The time evolution}
\label{sec: time_evolution}
As mentioned in Sec. \ref{sec: intro}, time does not exist a priori, and we may define it as follows.
We choose a basis, where $A_0$ is diagonal and its eigenvalues are in ascending order:
\begin{align}
A_0=\text{diag}(\alpha_1,\alpha_2,\ldots,\alpha_N)\, ,\quad \alpha_1\le\alpha_2\le\ldots\le\alpha_N \ .
\end{align}
Then, we define $\bar{\alpha}_k$ as
\begin{equation}
\bar{\alpha}_k = \frac{1}{n} \sum_{i=1}^{n} \alpha_{k+i}\ ,
\end{equation} 
and the time $t_\rho$ as
\begin{equation}
\label{eq: time}
t_\rho = \sum_{k=1}^{\rho} \abs{\bar{\alpha}_{k+1} -  \bar{\alpha}_k } \ .
\end{equation}
Here, we introduce the $n \times n$ matrices $\bar{A}_i(t)$ as
\begin{equation}
\qty(\bar{A}_i)_{ab}(t) = \qty(A_i)_{k+a, k+b} \ ,
\end{equation}
which represent space at time $t$.

After deforming the integration contour of the eigenvalues $\alpha_a$, we introduce the constraint $\alpha_N=\sqrt{\kappa}\in \mathbb{C}$ in order to obtain real time in the Lorentzian model.
This constraint is implemented numerically by adding a term $\gamma_\alpha (\alpha_N - \sqrt{\kappa})^4/4$ to the effective action and taking $\gamma_\alpha$ sufficiently large.
Note that once we introduce the above constraint, Eqs. (\ref{eq: TrA0sq}) and (\ref{eq: TrAisq}) do not hold anymore.

\section{Complex Langevin method}
The complex Langevin method (CLM) \cite{Parisi:1983mgm, Klauder:1983sp} is a stochastic process that can be applied successfully to many systems with a complex action problem.
One writes down stochastic differential equations for the complexified degrees of freedom, which can be used to compute expectation values under certain conditions.
Consider a model given by the partition function
\begin{equation}
Z=\int dx\, w(x) \ ,
\end{equation}
where $x\in \mathbb{R}^n$ and $w(x)$ is a complex-valued function.
In the CLM, we complexify the variables
\begin{equation}
x\in \mathbb{R}^n \longrightarrow z \in \mathbb{C}^n\ ,
\end{equation}
and solve the complex Langevin equation
\begin{equation}
\label{Langevin_eq}
\frac{dz_k}{d\sigma}={\frac{1}{w(z)}\pdv{w(z)}{z_k}} +{\eta_k(\sigma)}\ ,
\end{equation}
where $\sigma$ is the Langevin time.
The first term of the R. H. S of Eq. \eqref{Langevin_eq} is the drift term, and the second one is real Gaussian noise with probability distribution
\begin{equation}
\mathrm{P}(\eta_k(\sigma)) 
\propto \exp \qty(-\frac{1}{4} \int d\sigma \sum_k [\eta_k(\sigma)]^2)\ .
\end{equation}
In order to ensure that the CLM will give correct solutions, we apply the criterion that the drift term should be exponentially suppressed for large values \cite{Nagata:2016vkn}.

\subsection{Application of the CLM to the type IIB matrix model}
To apply the CLM to the type IIB matrix model, we a make change of variables \cite{Nishimura:2019qal}: 
\begin{equation}
\alpha_1=0\ ,\quad \alpha_i = \sum_{k=1}^{i-1} e^{\tau_k} \text{\quad (for $2 \leq i \leq N$)}\ ,
\end{equation}
where we introduce new real variables $\tau_k$. In this way, the ordering of $\alpha_i$ is automatically realized.
Initially, $\alpha_i$ are real, and $A_i$ are Hermitian matrices. To apply the CLM, we complexify  $\tau_k$ and take $A_i$ to be SL($N,\mathbb{C}$) matrices.
The CLM equations are given by
\begin{align}
\frac{d\tau_k}{d\sigma}
&=-{\pdv{S_\mathrm{eff}}{\tau_k}} +{\eta_k(\sigma)} \ ,\\
\frac{d(A_i)_{kl}}{d\sigma}
&=-{\pdv{S_\mathrm{eff}}{(A_i)_{lk}}} +{(\eta_i)_{kl}(\sigma)}\ ,
\end{align}
where $S_\mathrm{eff}$ is obtained from $\tilde{S}_\mathrm{b}$ in Eq. (\ref{k1}),  by adding the appropriate gauge fixing and change of variable terms.

In our simulations, we set $\beta=1$ and add the term
\begin{equation}
\frac{\gamma_\alpha}{4}(\alpha_N-\sqrt{\kappa})^4\, ,
\end{equation}
in order to enforce the constraint $\alpha_N = \sqrt{\kappa}$, where $\kappa\in \mathbb{C}$.

\section{Results}

\subsection{Expectation value of the time coordinate}
\begin{figure}
\centering
\includegraphics[scale=0.5]{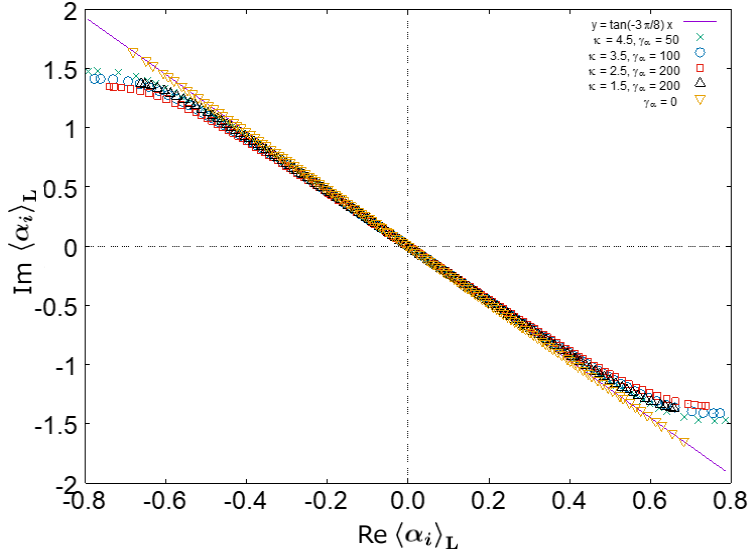} 
\caption{Expectation values of the  eigenvalues $\alpha_i$ of $A_0$ for $N=128$. The
$\gamma_\alpha=0$ case (inverted triangles) corresponds to the Euclidean model, where the complex phase of the expectation values $\expval{\alpha_i}_\mathrm{L}$ is $\exp(-i3\pi/8)$.
Those points lie, as expected, on the purple line.
When $(\kappa, \gamma_\alpha)=(4.5,50),(3.5,100),(2.5,200),(1.5,200)$, we obtain the curves that go through the circles, the squares, and the triangles, respectively.
The points near zero lie on the purple line, which implies that the emergent time is Euclidean.
On the other hand, as we move further from zero, we obtain almost real $(\Delta \alpha_i)_\mathrm{L}$, which implies the emergence of Lorentzian time.
}
\label{fig: alpha}
\end{figure}

When $\gamma_\alpha=0$, Eq. \eqref{eq: A0} holds, and we expect that
\begin{equation}
\expval{\alpha_i}_\mathrm{L} =e^{-i\frac{3\pi}{8}}\expval{\tilde{\alpha}_i}_\mathrm{E}\, .
\end{equation}
This is true because the Euclidean and the Lorentzian models are equivalent, and time corresponds to the Euclidean time. We measure the time differences
\begin{equation}
(\Delta \alpha_i)_\mathrm{L} 
= (\alpha_{i+1})_\mathrm{L} -(\alpha_i)_\mathrm{L}\ .
\end{equation}
If       $(\Delta \alpha_i)_\mathrm{L} \propto \exp(-i3\pi/8)$, the emergent time is Euclidean, 
while if $(\Delta \alpha_i)_\mathrm{L}\in\mathbb{R}$, the Lorentzian model gives real time with Lorentzian signature.

In Fig. \ref{fig: alpha}, we plot the expectation values of the time coordinates $\expval{\alpha_i}_\mathrm{L}$ on the complex plane.
Different symbols denote different sets of parameters $(\kappa, \gamma_\alpha)$.
The $\expval{\alpha_i}_\mathrm{L}$ close to the origin lie on the $\exp(-i 3 \pi/8)$ line, which implies that the emergent time is Euclidean.
On the other hand, as we move further from the origin, some of $\expval{(\Delta \alpha_i)_\mathrm{L}}$ are almost real and the emergent time is Lorentzian.

The metric signature change occurs dynamically.

\subsection{Time evolution of space}
The time evolution of the extent of space is given by
\begin{equation}
R^2(t)
=\expval{\frac{1}{n} \tr \qty(\bar{A}_i(t))^2}=e^{2i\theta}\abs{R^2(t)}\ .
\end{equation}
The matrices  $\bar{A}_i$ are complex; therefore $R^2(t)$ is also complex.
The time $t$ is defined in Eq. \eqref{eq: time}.
From Eq. (\ref{eq: TrAisq}) we see that when $\theta \sim \pi/8$, we obtain Euclidean space, and when  $\theta \sim 0$, we obtain real space from the Lorentzian model.
Therefore, the signature of spacetime can change dynamically in this model.

\begin{figure}
\centering
\includegraphics[scale=0.5]{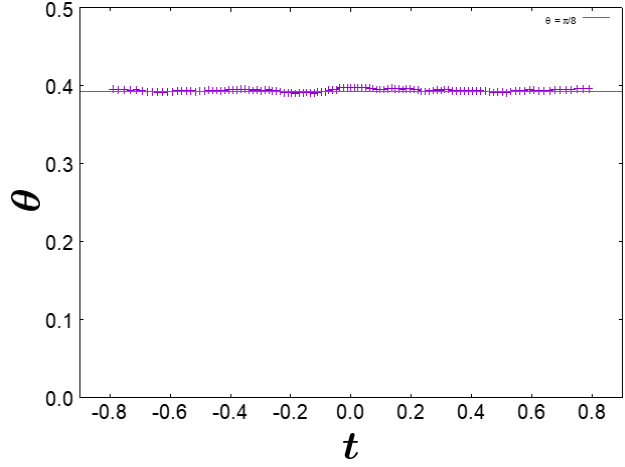} 
\caption{$\theta(t)$ is plotted against $t$ for $N=128$ and $(\kappa,\gamma_\alpha)=(4.5,50)$.
All values of $\theta(t)$ are on the $\theta=\pi/8$ line, which implies that the emergent space is Euclidean.}
\label{fig: theta}
\end{figure}
\begin{figure}
\centering
\includegraphics[scale=0.5]{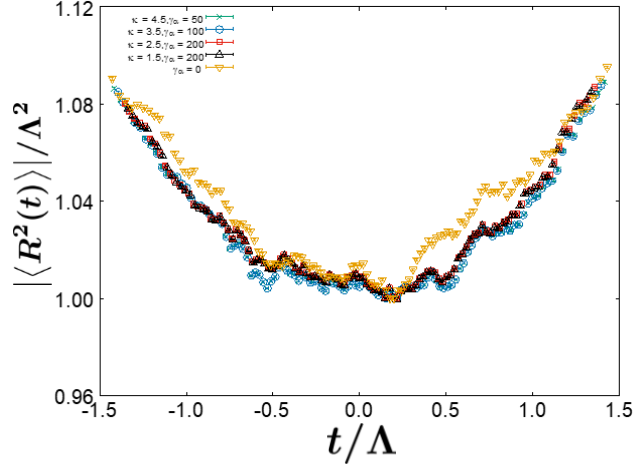} 
\caption{$|\expval{R^2(t)}|/\Lambda^2$ is plotted against $t/\Lambda$ for $N=128$ and $(\kappa, \gamma_\alpha)=(4.5,50)$, (3.5,100), (2.5,200), (1.5,200), where $\Lambda=\abs{\expval{R^2(0)}}^{1/2}$.
The symbols are the same as in Fig. \ref{fig: alpha}. We observe scaling behavior and a slight expansion of space as $t/\Lambda$ moves away from zero. 
}
\label{fig: Rsq}
\end{figure}
In Fig. \ref{fig: theta}, $\theta(t)$ is plotted against $t$ for $N=128$ and $(\kappa,\gamma_\alpha)=(4.5,50)$.
The green line represents $\theta=\pi/8$, and all values of $\theta(t)$ are on this line.
Therefore, space is the same as the one obtained by the Euclidean model for all times.

In Fig. \ref{fig: Rsq}, $|\expval{R^2(t)}|/\Lambda^2$ is plotted against $t/\Lambda$ for $N=128$ and various sets of parameters $(\kappa,\gamma_\alpha)$, where $\Lambda=\abs{\expval{R^2(0)}}^{1/2}$.
We observe a nice scaling behavior of the plots for all the parameters chosen. Furthermore, we observe a slight expansion of the extent of space with time, as we move away from the origin.

\section{Conclusions}
In this work, we successfully applied the CLM to the bosonic type IIB matrix model to overcome the sign problem.
This was made possible by Wick--rotating the model using a parameter $0\leq u \leq 1$, which interpolates smoothly between the Euclidean ($u=1$) and the Lorentzian ($u=0$) models.
These simulations avoid the approximation used in \cite{Kim:2011cr}, which yields a singular spacetime \cite{Aoki:2019tby}.
We find that the model is in a new phase, where smooth time emerges from the dynamics of the model and (noncommutative) space is continuous.
The $u=0$ model was simulated for the first time, and results were obtained without a $u\to 0$ extrapolation.
We showed that the Lorentzian and the Euclidean models are equivalent, and expectation values can be analytically continued from one model to the other.
The expectation values (\ref{eq: TrA0sq}) and (\ref{eq: TrAisq}) in the Lorentzian model are complex, and spacetime is Euclidean.

We studied a scenario for the dynamical emergence of the Lorentzian signature by introducing a constraint $\alpha_N=\sqrt{\kappa}\in \mathbb{C}$.
Then, the Euclidean and the Lorentzian models are not equivalent anymore, and time in the Lorentzian model, arising from the expectation values of $A_0$, may turn out to be real.
We defined time from the differences $(\Delta \alpha_i)_\mathrm{L}$ and showed that, although complex near the origin, they turn out to be real at later times.
This provides the context for a scenario where the signature of the metric changes {\it dynamically} from Euclidean to Lorentzian. 

We also studied the evolution of the extent of space with time. In this model, space turns out to be Euclidean for all times and exhibits a slight expanding behavior with time. 

We expect that supersymmetry will play an essential role in obtaining a phenomenologically viable theory. We are currently investigating its effect, which we will report in a future
publication. 

\acknowledgments
T.~A., K.~H., and A.~T. were supported in part by Grant-in-Aid (Nos. 17K05425, 19J10002, and 18K03614, 21K03532, respectively) from Japan Society for the Promotion of Science.
This research was supported by MEXT as ``Program for Promoting Researches on the Supercomputer Fugaku'' (Simulation for basic science: from fundamental laws of particles to creation of nuclei, JPMXP1020200105) and JICFuS. 
This work used computational resources of supercomputer Fugaku provided by the RIKEN Center for Computational Science (Project ID: hp210165) and Oakbridge-CX provided by the University of Tokyo (Project IDs: hp200106, hp200130, hp210094) through the HPCI System Research Project . 
Numerical computation was also carried out on PC cluster in KEK Computing Research Center.
This work was also supported by computational time granted by the Greek Research and Technology Network (GRNET) in the National HPC facility ARIS, under the project IDs SUSYMM and SUSYMM2.

\bibliography{ref_CLMIKKT}

\end{document}